\documentclass[12pt] {article}   
\usepackage{amssymb,amsmath}
\usepackage[round]{natbib} 
\usepackage{color}
\usepackage{hyperref}
\usepackage{graphicx}

\newcommand{\bal}[1]{  \begin{align} \label{#1} }  
\newcommand{\beq}[1]{  \begin{equation} \label{#1} }  
\newcommand{\eeq}{     \end{equation}}  
\newcommand{\beqa}[1]{ \begin{eqnarray} \label{#1} }	
\newcommand{\eeqa}{\end{eqnarray}  }

\renewcommand{\appendix}{
  \setcounter{section}{0}\renewcommand{\thesection}{\Alph{section}}
  \section*{Appendix} 
}

\def\Appendix#1{
  \setcounter{equation}{0}
  \renewcommand{\theequation}{\thesection.\arabic{equation}}
  \section{#1}
}

\newtheorem{thm}{Theorem}
\newtheorem{lem}{Lemma}

\newcommand{\rf}[1]{(\ref{#1})}

\def\bd#1{\mbox{\boldmath$\displaystyle\mathbf{#1}$} }
\def\tens#1{\mathbb{\,#1}}	   
\def\tr{\operatorname{tr}} 
 
\def\det{\operatorname{det}}

\def\singlespacing{\baselineskip=13pt}	

\def\rev#1{#1}	   

\begin{document} 

\singlespacing

\title{
\textcolor{blue}{Invariants of ${\bd C}^{1/2}$ in terms of the invariants of ${\bd C}$}  
}

\author{Andrew N. Norris\\ \\    Mechanical and Aerospace Engineering, 
	Rutgers University, \\ Piscataway NJ 08854-8058, USA \,\, norris@rutgers.edu 	
}	\date{}

\maketitle

\begin{abstract}
The three invariants of ${\bd C}^{1/2}$ are key to expressing this tensor and its inverse  as a polynomial in  ${\bd C}$.  Simple and symmetric expressions  are presented  connecting  the two sets of invariants  $\{I_1, I_2,I_3\}$ and $\{i_1, i_2,i_3 \}$ of ${\bd C}$ and ${\bd C}^{1/2}$, respectively.  The first result is a bivariate  function    relating   $I_1, I_2$ to  $i_1, i_2$.   The functional form of $i_1$  is the same as that of  $i_2$ when the roles of the  ${\bd C}$-invariants are reversed.  The second result expresses the invariants using a single function call.  The two sets of expressions emphasize symmetries in the relations among these four invariants.

\end{abstract}

\section{Introduction}\label{sec1}

We consider relations among the basic tensors of three dimensional continuum
mechanics, all defined by the deformation $\bd F$,  
\beq{1}
{\bd F} = {\bd R}{\bd U}= {\bd V}{\bd R}, 
\qquad {\bd C}  = {\bd F}^t{\bd F},
\qquad {\bd B}  = {\bd F}{\bd F}^t. 
\nonumber
\eeq
 ${\bd U}$ and ${\bd V}$ are are symmetric  and  positive definite,  and therefore
\beq{2292}
{\bd U}  = {\bd C}^{1/2},
\qquad
{\bd V}  = {\bd B}^{1/2}. \nonumber
\eeq
Here we will only consider properties of $\bd U$ and $\bd C$, but the results apply  to ${\bd V}$ and $\bd B$. 

Although the square root of a second order positive definite symmetric tensor is unique and unambiguous it is not, however, a simple algebraic construct.  One way to circumvent this problem is to express ${\bd U}$ as a polynomial in ${\bd C}$ using the Cayley-Hamilton equation,  
\beq{201}
{\bd U}^3 - i_1{\bd U}^2 + i_2{\bd U}  - i_3 {\bd I} =0.  
\eeq 
Here $i_1, i_2, i_3$ are the   invariants  of $\bd U$, 
\beq{3}
 i_1  = \tr {\bd U}, \qquad
   i_2  =  \frac12  (\tr {\bd U})^2 - \frac12  \tr {\bd U}^2 ,
   \qquad
  i_3   = \det {\bd U} ,
  \nonumber
\eeq
Multiply \rf{201} by $({\bd U}  + i_1 {\bd I})$, and note that the result contains terms proportional to $\bd I$, 
${\bd U}$, ${\bd U}^2$ and ${\bd U}^4$.   Replacing the latter two by 
${\bd C}$ and ${\bd C}^2$ gives  \citep{Ting1985}
\beq{202}
{\bd U}  =  ( i_1 i_2 - i_3)^{-1} \big( i_1i_3{\bd I}  + ( i_1^2 - i_2){\bd C} - {\bd C}^2\big).     
\eeq
Note that $i_1 i_2 - i_3 = \det( i_1 {\bd I} - {\bd U}) > 0$ \citep{Carroll04}.  The inverse ${\bd U}^{-1}$ may be obtained by multiplying each side of \rf{202} with ${\bd C}^{-1}$ and  using the Cayley-Hamilton equation for ${\bd C} $ to eliminate the single remaining ${\bd C}^{-1}$ term.  The orthogonal rotation tensor  follows as ${\bd R} = {\bd F}{\bd U}^{-1}$, from which one can determine kinematic quantities such as 
  the rotation angle and the axis of rotation \citep{Guansuo98}. 

The relation \rf{202} for ${\bd U}$  in terms of ${\bd C}$  avoids the tensor square root difficulty  but introduces another:  how to express  $\{i_1,i_2,i_3\}$ in terms of ${\bd C}$, or more specifically in terms of its  invariants,   
\beq{31}
 I_1  = \tr {\bd C}, \qquad
   I_2  =  \frac12  (\tr {\bd C})^2 - \frac12  \tr {\bd C}^2 ,
   \qquad
  I_3   = \det {\bd C} .   \nonumber
\eeq
While the relation $i_3 = \sqrt{I_3}$ is simple, formulas for $i_1$ and $i_2$ are not.
But as eq. \rf{202} and related identities illustrate,  the  functional relations between the two sets of invariants are important for obtaining semi-explicit expressions for stretch and rotation tensors,  and for their  derivatives \citep{Hoger84b,Steigmann02,Carroll04}. 

 The first such relations are due to \citet{Hoger84} who derived   expressions for $\{i_1,i_2\}$    by solving  a quartic equation. \citet{Sawyers86} subsequently showed that one can obtain alternative relations using the standard solutions \citep{Goddard97}  for the cubic equation of the eigenvalue of $\bd C$.   Let  $ \lambda_1,  \lambda_2,  \lambda_3$ be the (necessarily positive) eigenvalues of $\bd U$, then 
the  eigenvalues of $\bd C$ are $ \lambda_1^2,  \lambda_2^2,  \lambda_3^2$, and 
\begin{subequations}\label{6}
\bal{6a}
&i_1 = \lambda_1 +\lambda_2  +\lambda_3 ,
 \qquad 
   i_2 = \lambda_1  \lambda_2 +\lambda_2  \lambda_3 +\lambda_3 \lambda_1 ,
   \qquad
  i_3 = \lambda_1  \lambda_2  \lambda_3 ,  
\\
&I_1 = \lambda_1^2 +\lambda_2^2 +\lambda_3^2,
 \qquad 
   I_2 = \lambda_1^2 \lambda_2^2+\lambda_2^2 \lambda_3^2+\lambda_3^2\lambda_1^2,
   \qquad
  I_3 = \lambda_1^2 \lambda_2^2 \lambda_3^2.  
  \label{6b}
\end{align}
\end{subequations}
 Sawyers' approach is to essentially compute the eigenvalues of  $\bd C$, take their square roots and from these determine the invariants of $\bd U$ by \rf{6a}. \citet{Jog02}  generalized this scheme to tensors of order $n>3$.   This method  does not  provide direct relations between the invariants.   
Although the formulas of \citet{Hoger84} and of \citet{Sawyers86} are explicit,   they are not totally  satisfactory.   In each case the functional forms are complicated.  As we will see, there is no way to avoid this ``complexity" since we are dealing with roots of cubic and quartic equations.  But that is not the basic issue, rather it  is a lack of any underlying symmetry or balance in the solutions of  \citet{Hoger84} and of \citet{Sawyers86}. This  makes it difficult to comprehend the formulas, and to place them in context.  It is all the more unsettling by virtue of the fact that the formulas are associated with algebraic systems of deformation tensors, systems that are elegant and generally quite transparent.  

The object of this paper is to express $\{i_1,i_2,i_3\}$ in terms of $\{I_1,I_2,I_3\}$ in two  forms that each display the underlying symmetry of the relations.   Both  forms employ a single function, but have slightly different properties.  
We begin in section \ref{sec2} with a summary of the principal  results, followed by a review of the previously known solutions in section \ref{sec3}.   The new formulas for the invariants of ${\bd C}^{1/2}$ are derived in section \ref{sec4}, with some closing comments in section \ref{sec5}.

\section{Principal results}\label{sec2}

The first result is 
\begin{thm}\label{thm1}
The invariants of ${\bd C}^{1/2}$ are 
\beq{000}
i_1 = I_3^{1/6} f\big( \frac{I_1}{I_3^{1/3}},  \frac{I_2}{I_3^{2/3}}\big), 
\qquad
i_2 = I_3^{1/3} f\big(   \frac{I_2}{I_3^{2/3}}, \frac{I_1}{I_3^{1/3}}\big), 
\qquad
i_3 = I_3^{1/2},
\eeq
where $f$ is a function of two variables, 
\begin{subequations}\label{001}
\bal{001a}
f(x,y) &= g(x,y)+ \sqrt{x-g^2(x,y)+ 2/g(x,y)},
\\
g(x,y) &=  \bigg( \frac13 \big(x + \sqrt{x^2 -3y}\big[ (\zeta+\sqrt{\zeta^2-1} )^{1/3}
+ (\zeta-\sqrt{\zeta^2-1} )^{1/3} \big] \big) \bigg)^{1/2}, 
 \\
 \zeta &=    \frac{27+ 2 x^3- 9 xy}{2( x^2 -3y)^{3/2} }. \label{c}
 \end{align}
 \end{subequations}
\end{thm}

The function  $g$ can be expressed in the alternate form 
\beq{090}g(x,y) =   \sqrt{ \frac13\big(x + 2\sqrt{ x^2 -3y } \, \cos (\frac{1}{3}\arccos \zeta (x,y))  \big)}. \nonumber
\eeq

It is clear from Theorem \ref{thm1} that the following reduced quantities are the important variables: 
\begin{subequations}\label{011}
\bal{011a}
j_1 &= \frac{i_1}{i_3^{1/3}},\qquad 
j_2 = \frac{i_2}{i_3^{2/3}}, 
\\
J_1 &= \frac{I_1}{I_3^{1/3}},\qquad 
J_2 = \frac{I_2}{I_3^{2/3}},   
\end{align}
\end{subequations}
in terms of which the Theorem states
\beq{77}
j_1 = f(J_1,J_2),\qquad j_2 = f(J_2,J_1).  \nonumber
\eeq

Alternatively,  the sum and difference of reduced invariants may be considered as  the key parameters, which is evident from: 
\begin{lem}\label{lem1}
The following  relation holds between the invariants of ${\bd C}$ and ${\bd C}^{1/2}$:
\beq{222}
\frac{J_1-J_2}{j_1-j_2} = j_1+j_2 + 2. 
\eeq  
\end{lem}
An immediate consequence of Lemma \ref{lem1} is that we need only determine $j_1+j_2$ or $j_1-j_2$ since the other follows directly from \rf{222}.  For instance, we could calculate 
$j_1+j_2 = f(J_1,J_2) +  f(J_2,J_1)$, but  this  requires evaluation of $f$ twice, and it does not reveal the underlying symmetry of the arguments. 
The second result is a simpler relation between the invariants, one that uses a single call to the function $f$: 
 \begin{thm}\label{thm2} 
 The reduced invariants of ${\bd C}^{1/2}$ and  ${\bd C}$ are connected by 
 \begin{subequations}\label{043}
\bal{043a}
j_1 &= \frac{s}{2}  + \frac{J_1-J_2}{2s+4}, 
\\
j_2 &= \frac{s}{2}   - \frac{J_1-J_2}{2s+4},  
\\
s  &=F(J_1,J_2),  
\end{align}
\end{subequations}
where 
\beq{088}
F(x,y) =  (2+x+y)^{1/3} \, f\big(\frac{6+x+y}{(2+x+y)^{2/3}}, \, \frac{9+5x+5y+xy}{(2+x+y)^{4/3}} \big) . 
\eeq
\end{thm}
Thus,  $j_1+j_2 = F(J_1,J_2)$ and  $F$ is  a  symmetric function   of its arguments, $F(x,y) =   F(y,x)$.  Also, $F$ provides an 
alternative expression  for the function $f$:
\beq{007}
f(x,y)= \frac12 F(x,y)  + \frac{x-y}{2F(x,y)+4}.  \nonumber
\eeq
This form for $f$ employs the function itself, but evaluated at different arguments.  This is a property of the  nonlinear nature of the function. 

\section{The methods of Hoger and Carlson and of Sawyers  }\label{sec3}

Starting from the identities \rf{6} it may be easily verified that \citep{Hoger84}
\beq{62}
   i_1^2-2i_2= I_1,
\qquad
 i_2^2 - 2i_1 i_3= I_2, 
\qquad
  i_3^2= I_3. 
\eeq
The last implies $i_3 = I_3^{1/2}$. 
It remains to find $i_1$ and $i_2$.  

\citet{Hoger84} eliminated $i_2$ between eqs. $\rf{62}_1$ and $\rf{62}_2$, to obtain a quartic equation in $i_i$ which they then solved.   The same solution for $i_i$ is obtained more directly by starting with the \emph{ansatz}
\beq{8}
i_i = \lambda  + \rho, 
\eeq
where $\lambda $ is any one of the triplet $\{\lambda_1 ,  \lambda_2 ,  \lambda_3\}$. For instance, if $\lambda = \lambda_1$ then $\rho =  \lambda_2 +  \lambda_3$ and $i_2= \lambda_1 ( \lambda_2 +\lambda_3 )+\lambda_2  \lambda_3 $ is 
\beq{9}
i_2    =   \rho \lambda + i_3/\lambda  . 
\eeq
This holds no matter which value $\lambda $ takes.  Substituting from \rf{8} into $\rf{62}_1$ implies
\beq{10}
\rho^2 = I_1 - \lambda^2+ 2i_3/\lambda . 
\eeq
The right member is necessarily positive, and using $i_3 = I_3^{1/2}$ we can therefore express $\rho >0$ in terms of $I_1$, $I_3$ and $\lambda $.  

In summary, 
\begin{subequations}\label{13}
\bal{13a}
i_1 = &\lambda   +\sqrt{I_1-\lambda^2 +2 \sqrt{I_3} /\lambda   } ,
\\
i_2=  &\sqrt{I_3}/\lambda   +\sqrt{I_1\lambda^2 - \lambda^4 +2 \sqrt{I_3} \lambda  } ,
\label{13b}
\\
i_3= &\sqrt{I_3},
\end{align}
\end{subequations}
where $\lambda$ is any positive root of the characteristic  equation of $\bd C$, 
\beq{55}
\lambda^6- I_1\lambda^4 +I_2\lambda^2 - I_3=0. 
\eeq
For instance, 
\beq{20}
 \lambda  = \bigg( \frac13\big(I_1 + \big[\xi+\sqrt{\xi^2-( I_1^2 -3I_2)^3} \big]^{1/3}
 																+ \big[\xi-\sqrt{\xi^2-( I_1^2 -3I_2)^3} \big]^{1/3}  \big) \bigg)^{1/2}, \nonumber
 \eeq
 and 
 \beq{21}
 \xi =    \frac12 (2 I_1^3- 9 I_1I_2+27 I_3).  \nonumber
 \eeq
\rev{Note that we assumed that $\lambda$ in the \emph{ansatz} \rf{8} and \rf{9} is a root of eq. \rf{55}, but this is actually a requirement, as can be seen from eqs. $\rf{62}_1$ using \rf{8} - \rf{10}.  Equations \rf{8} and \rf{9} represent a standard method of reducing a quartic to a cubic equation. }
 
Equation \rf{13a} is essentially the same as the first relation\footnote{It should be noted that the second relation  in their eq. (5.5) never applies.  That is, it can be shown that the possible equality cannot occur.} 
 \cite[eq. (5.5)]{Hoger84}, although they did not identify  the root of the cubic explicitly.  \citet{Hoger84} recommended  using $\rf{62}_2$ to obtain $i_2$.  The relation \rf{13b} is quite different and is suggestive of the symmetry underlying  the solutions for $i_1$ and $i_2$ that is evident in Theorem \ref{thm1}. We discuss this further in the next section from a different perspective.  
 
 It is interesting to compare this with the  explicit positive solution  of \rf{55}  provided by \cite{Guansuo98}, based on    \citet{Sawyers86}.   Starting with the characteristic equation for $\bd U$, 
 \beq{56}
\lambda^3- i_1\lambda^2 +i_2\lambda  - i_3=0, 
\eeq
combined with eqs. $\rf{62}_2$ and $\rf{62}_3$ , this becomes a quadratic equation for $i_2$. 
The solution is \cite[p. 199]{Guansuo98} 
 \beq{3073}
 i_2 = \lambda^{-1} \big( \sqrt{I_3} + \sqrt{2\sqrt{I_3} \lambda^3 +I_2\lambda^2 -I_3} \big) .  
 \nonumber
 \eeq
 This appears to be different than  eq. \rf{13b}, but they are equivalent when one takes into account that $\lambda$ satisfies eq. \rf{55}.  

In short, \citet{Hoger84} and \citet{Sawyers86} derived eqs. \rf{13a} and \rf{13b}, respectively.  They did not however, note the symmetry between the formulas, which is one of the central themes in this paper: that a single function determines both $i_1$ and $i_2$.   In the next section we complete the proof of Theorem \ref{thm1}.

\section{An alternative approach}\label{sec4}

The three conditions in \rf{62} can be combined into a single polynomial identity, 
\beq{6121}
(1-i_2z^2)^2 + (i_1z-i_3z^3)^2 = 1+I_1z^2+I_2z^4+I_3z^6, \quad \forall \, z \in \tens{C}.  
\nonumber
\eeq
Using the reduced variables of \rf{011},  this becomes 
\beq{012}
(1-j_2z^2)^2 + (j_1z- z^3)^2 = 1+J_1z^2+J_2z^4+ z^6, \quad \forall \, z \in \tens{C}.  
\nonumber
\eeq

Comparing coefficients implies the pair of coupled equations
\begin{subequations}\label{044}
\bal{044a}
j_1^2-2j_2 &= J_1,  \\
j_2^2-2j_1 &= J_2.
\end{align}
\end{subequations}
Thus,  solutions must be of the form
\beq{014}
j_1 = f(J_1, J_2),\qquad j_2 = f(J_2, J_1), \nonumber 
\eeq
for some function $f(x,y)$ which satisfies
\beq{015}
f^2(x,y) - 2f(y,x) - x = 0. 
\eeq
This is the fundamental equation for $f(x,y)$.  It implies the dual relation
\beq{0151}
f^2(y,x) - 2f(x,y) - y = 0. 
\eeq
Eliminating $f(y,x)$ between eqs. \rf{015} and \rf{0151} leads to a  quartic in $f= f(x,y)$: 
\beq{016}
(f^2 - x)^2-8f-4y=0.
\eeq

This  is equivalent to the quartic of \cite{Hoger84} but expressed in the reduced variables.  We have already 
derived a solution of the quartic in the previous section by using the \emph{ansatz} \rf{8} based on a root of the cubic equation  \rf{55}. 
Equation \rf{13a} therefore \emph{defines} the function $f$, which can be read off by converting to the reduced variables $j_1, j_2,J_1, J_2$.  It may be easily verified that the function of eq. \rf{001} results. 

But what about the relation \rf{13b} for $i_2$?  It does not seem to convert into the expression claimed in Theorem \ref{thm1}, i.e. $j_2 = f(J_2,J_1)$.  Rather, using eq. \rf{13b} and $j_2 = f(J_2,J_1)$ to define $f$ we obtain a different  expression for $f$: 
\beq{005}
f(y,x) = \frac{1}{g(x,y)}+ \sqrt{x-g^2(x,y)+ 2/g(x,y)}\, g(x,y). 
\eeq
This is in fact  consistent with the definition of $f$ in Theorem \ref{thm1} because   $g(x,y)$ satisfies the normalized version of eq. \rf{56}, 
\beq{004}
 g^3(x,y) - f(x,y)g^2(x,y) +f(y,x)g(x,y) - 1= 0. 
\eeq
Using this and the expression for $f(x,y)$ in \rf{001}, gives eq. \rf{005}.   This completes the proof of Theorem \ref{thm1}.

It is interesting to note from \rf{004} that $1/g(y,x)$ satisfies the same equations as $g(x,y)$, i.e. 
\beq{006}
 g^{-3}(y,x) - f(x,y)g^{-2}(y,x) +f(y,x)g^{-1}(y,x) - 1= 0.  \nonumber
\eeq
But this does not mean that $g(y,x) $ equals $1/g(x,y)$, since they can (and do) correspond to different roots of the cubic. 

The identity \rf{222} of Lemma \rf{lem1} follows from the coupled equations \rf{044}, and the details of the proof of Theorem \ref{thm2} are in the appendix.

\section{Conclusion}\label{sec5}

Although the expressions for $i_1$ and $i_2$ involve the roots of the characteristic cubic equation of $\bd C$, it seems that the governing  quartic equation \rf{016}  is more fundamental. This is the equation that defines the functions $f$ and $F$ of Theorems \ref{thm1} and   \ref{thm2}.  In fact $F$ is defined by $f$, which is in some ways the central function involved.  
It is interesting that the quartic equation first considered by \citet{Hoger84} reappears in this manner. 

Which of the expressions for $i_1$ and $i_2$ are actually best in practice?  While the expressions \rf{043} are perhaps the most aesthetically pleasing in form,  eq. \rf{000} is probably simpler to implement.  The final choice is of course left to the reader. 

 \appendix
\Appendix{Proof of Theorem \ref{thm2}}

For simplicity of notation, let $f$ and $f'$ denote $f(x,y)$ and $f(y,x)$, respectively.  Then the coupled equations 
\rf{015} and \rf{0151} are 
\begin{subequations}
\begin{align}
f^2 - 2f' & =x, \nonumber
\\
{f'}^2-2f &= y. \nonumber
\end{align}
\end{subequations}
Adding and subtracting  yields, respectively, 
\begin{subequations}
\begin{align}
(f+f'-1)^2 & = 1+x+y + 2ff', \nonumber
\\
(f-f')(f+f'+2)&= x-y, \nonumber
\end{align}
\end{subequations}
which in turn imply
\begin{subequations}\label{503}
\bal{503a}
f+f'& = F,
\\
f-f' &= \frac{x-y}{F+2}, 
\end{align}
\end{subequations}
where
\beq{504}
F = 1 + \sqrt{1+x+y + 2ff'}.  \nonumber
\eeq
The function $F= F(x,y)$ is clearly a symmetric function of $x$ and $y$, i.e., it is unchanged if the arguments are switched. 
Solving eqs. \rf{503} gives
\begin{subequations}\label{505}
\bal{505a}
f& = \frac{F}{2} + \frac{x-y}{2F+4},
\\
f' &= \frac{F}{2} - \frac{x-y}{2F+4}. 
\end{align}
\end{subequations}

Although the formulas \rf{505} clearly split $f$ into parts that are symmetric and asymmetric in the two arguments, they are not explicit since the function $F$ involves the product $ff'$.  Taking the product of the two expressions in \rf{505} leads to an equation for $ff'$.  It is simpler to consider the equation for $F$, which after some manipulation may be reduced to the  quartic: 
\beq{588}
[F^2-(6+x+y)]^2- 8(2+x+y)F  - 4[ (5+x)(5+y)-16]=0.  \nonumber
\eeq
Let
\beq{589}
F =     (2+x+y)^{1/3} \,u, \nonumber
\eeq
then $u$ satisfies 
\beq{0161}
(u^2 - X)^2-8u-4Y=0, 
\eeq
where 
\beq{101}
X = \frac{6+x+y}{(2+x+y)^{2/3}}, \qquad Y = \frac{9+5x+5y+xy}{(2+x+y)^{4/3}}. \nonumber
\eeq 
Equation \rf{0161} is the same as the quartic \rf{016} satisfied by $f$, but with $X$ and $Y$ instead of $x$ and $y$.  Thus, 
\beq{017}
u = f(X,Y), \nonumber
\eeq
which completes the proof of Theorem \ref{thm2}. 


\end{document}